\documentclass[aps,jcp,showpacs,twocolumn,amssymb]{revtex4-2}
\usepackage{amssymb}   
\usepackage{graphicx}
\usepackage{graphics}
\usepackage{color,epsfig}
\usepackage{amsmath}
\usepackage{bm}        
\usepackage{multirow}
\usepackage{booktabs}
\usepackage{float}
\usepackage{fancyhdr}
\usepackage[latin1]{inputenc}

\begin{document}

\title{Universal behavior in complex-mediated reactions: Dynamics of
 S($^{1}D$)+ o-D$_2$ $\rightarrow$ D + SD at low collision 
energies}

\author{Manuel Lara\footnote{Corresponding author. E-mail:
 {\em manuel.lara@uam.es}}}
\affiliation{ Departamento de Qu\'imica F\'isica Aplicada, Facultad
de Ciencias, Universidad Aut\'onoma de Madrid, 28049 Madrid, Spain}

\author{P. G. Jambrina}
\affiliation{Departamento de Qu\'imica F\'isica, Facultad
de Farmacia, Universidad de Salamanca, 37008 Salamanca, Spain
Spain}

\author{F. J. Aoiz}
\affiliation{Departamento de Qu\'imica F\'isica, Facultad de Qu\'imica, Universidad
Complutense, 28040 Madrid, Spain}

\date{\today}

\begin{abstract}

Reactive and elastic cross-sections, and rate coefficients, have been
calculated for the S($^{1}D$)+ D$_2$($v$=0, $j$=0) reaction using a modified
hyperspherical quantum reactive scattering method. The considered collision
energy ranges from the ultracold regime, where only one partial wave is open,
up to the Langevin regime, where many of them contribute. This work  presents
the extension of the quantum calculations, which were compared with the
experimental results in a previous work, down to energies in the cold and
ultracold domains. Results are analyzed and compared with the universal case
of the quantum defect theory by Jachymski {\em et al.} [Phys. Rev.
Lett.\,{\bf 110}, 213202 (2013)].  State-to-state integral and differential
cross sections are also shown covering the ranges of low-thermal, cold and
ultracold collision energy regimes. It is found that at $E/k_BT<1$\,K  there
are substantial departures from the expected statistical behavior, and that
dynamical features become increasingly important with decreasing collision
energy, leading to vibrational excitation.

\end{abstract}

\maketitle

\section{Introduction}

Nowadays, state-of-the-art  experimental techniques allow the access to the
cold regime ($T< 1$\,K) with extraordinary resolution, and make it possible to
explore the mostly unknown ultracold region
($T<1$\,mK).\cite{Nar1,Nar2,Nar3,Nar4,Nar5,Osterwalder:EPJ15,Osterwalder:NC18,PDRKN:NC21,Plomp2021}.
Accordingly, the focus of the field of (ultra-)cold reaction dynamics is moving
from reactions only involving alkali atoms and dimers
\cite{Staanum,Zahzam,Wynar,Mukaiyama,Syassen,hudsonexp,Ospel} to more generic
bimolecular systems.
\cite{Canosa,nuestroFara,Costes,Costes2,FaraBas,Syncro,Zeem,2017_Science_Perreault,2018_NatChem_Perreault,sarp_hed2_science,Zhou:NC22}
The availability of new experimental results \cite{Yu_Liu:Nature_2021} is
calling for computational studies to be extended to the low energy regime.

Significant progress has been achieved in the theoretical treatment of the dynamics and
stereodynamics of  {\em inelastic} atom-diatom and diatom-diatom collisions, and in
the simulation of the associated experiments.\cite{2018_NC_sharples,2019_PRL_Jambrina,2023_PRL_Jambrina,Meeraker:S20,2019_NC_Heid,2022_JPCL_Jambrina}
However, quantum results for {\em reactive processes} at collision energies, $E\leq
1$\,K, should be taken with some reservations. Given the extreme sensitivity of the
scattering length to small details of the PES,  the results of ``exact'' quantum
calculations at ultracold energies can be always put into question.  As a general
principle, the uncertainty of the potential energy should be much lower than the
considered collision energy in order to accurately describe the outcome of a collision
process, and  errors in potential energy calculations of reactive systems at short range
(SR) are still typically high. In this scenario, and for generic light atom+diatom
reactions, quantum calculations for energies below $\sim 1$\,K may lack predictive
power. However, the dynamics of some particular systems, dubbed {\em universal},
show an interesting property: a weak dependence on the details of the PES at short
range that makes possible to go much further and to draw predictions at much lower
energies. In this regard, let us note that reactions which are accessible at cold
energies are generally  barrierless. Under these conditions, long-range (LR) interactions,
easier to calculate than SR ones,  will determine the amount of incoming flux which is
able to reach the transition  state region of the PES and thus to react.  In fact, as we get
closer in total energy to the reactants asymptote, and due to the existence of quantum
reflection, the influence of LR interactions becomes paramount, and may lead to the
{\em universal} behavior:~\cite{Fara} the outcome of the collision depends exclusively
on the analytical LR interactions and not on the SR (chemical) forces. The idea of
universality is formalized in the context of multi channel quantum-defect theory
(MQDT), allowing LR parametrizations that can be used to fit the experimental data.
\cite{Jach:02,Jach:03,Jach:04,Oster14,Nar5,Croft2020,Frye2015} In this case, the
information given by the Wigner laws,~\cite{Sade} together with the knowledge of a
few parameters such as the reduced mass and the $C_6$ or $C_4$ coefficients, may
completely determine the value of the reaction rate coefficient at low enough collision
energies.

These ideas are not new. Two widely used classical models in reaction dynamics: i) the
(Langevin) capture model,\cite{Lange} and ii) the statistical model, \cite{RGM:JCP03}
are based on an analogous assumption: the possibility of disregarding the inner part of
the PES. In the former model, the accent is put on the capture process: once the system
is captured by the $\sim R^{-n}$ LR forces, the reactive complex is formed and its fate is
irreversible: it will dissociate to form the products. As there is no way back, a detailed
description of the inner part or the PES is deemed unimportant, the only function of the
latter being to transfer the system from the reactants to the products valley. In the
statistical models, the capture process to form the complex, and its reverse (the
dissociation process which decomposes it) are the essential ingredients. LR forces lead
to form the complex and are the way to break it. Again, the exact details of the inner
PES are irrelevant, as long as the behavior is so ergodic that allows a description of the
correspondence between incoming and outgoing channels in terms of probabilities. In
short, one may disregard the details of the interaction at SR in both cases.

In our recent theoretical work, we have tried to explore and link these two different
models and their behavior in the ultracold limit. \cite{LaraPRA,LaraJCP2015}  In fact,
both models are not unrelated: a good way to assure the irreversibility of the capture
process in a complex-mediated system is to increase the number of possible
fragmentation channels starting from the complex.  Accordingly, when the number of
product channels overrides that of the reagents  a complex mediated reaction with
strong ergodicity can be considered a Langevin-type system: once captured in the
complex, there is no significant probability to form back the reagents.

In previous works,\cite{LaraPRA,LaraJCP2015} we devoted our attention to the
exothermic D$^{+}$ + {\em para}-H$_2$ $\rightarrow$ H$^+$ + HD reactive
collision at cold and ultracold energies. Time-independent quantum scattering
calculations were carried out  using a  modified hyperspherical quantum
reactive scattering method \cite{Launayfirst,Lara2}. Quantum results were
obtained for energies as low as $10^{-8}$\,K in this prototype insertion
reaction, whose $\sim R^{-4}$ LR behavior implies propagations up to very large
distances. To the best of our knowledge, no other work in the literature has
attained so low energies in an atom + molecule system with such an extended LR.
We analyzed the behavior found in the ultracold regime in terms of the
quantum-defect model in Ref.~\citenum{Jach:02}. The D$^{+}$ + {\em para}-H$_2$
system was presented as a good example of a {\em non-universal} system, since
the number of channels of the products is comparable to that of the reagents.

In this work,  we will extend a prior study of the S($^{1}D$)+ D$_2$($v$=0, $j$=0)
$\rightarrow$ SD + D process\cite{LaraJPC2016} in the energy range 1\,K--220\,K
(see Fig. 3 in Ref.~\citenum{LaraJPC2016}) to the cold and ultracold regimes. The SD$_{2}$
system constitutes a prototype of insertion reactions governed by an $R^{-6}$
LR potential and has attracted a great deal of attention from both theoreticians
and experimentalists.
Being mediated by a collision complex, statistical models, which will be relevant
in our discussion, have been applied to the system and its isotopic variants. 
They have been found to account very well for the QM results.\cite{RGM:JCP03,G-L:IRPC2007,AGS:JCP08,Guo:JCP2005}
A few years ago, excitation functions for the S($^{1}D$)
+ D$_2$($v$=0, $j$=0) $\rightarrow$ SD + D reaction in the near cold regime
were measured and published in coincidence with their theoretical counterparts.
\cite{LaraJPC2016} This experiment, performed using  the angle-variable crossed
molecular beam technique, complemented the set of measurements at unprecedented
low temperatures  carried out in Bordeaux since 2010 on the isotopic variants
of the SH$_2$ system.\cite{C0CP02705G,bert:10,PhysRevLett.109.133201,JLA:JCP21}
An overall good agreement was found between the experimental data and
theoretical calculations performed with different methods.
In this new article we analyze in depth the already published QM cross sections,
 and study for the first time the behavior of the system in the cold and ultracold
 regimes through the analysis of final-state resolved cross sections and reaction probabilities.
We show that in many respects it can be considered as
universal, and explore to what extent such assumption is valid. As for
the accuracy of the results, the same caveats as in similar works in the
ultracold regime hold here. \cite{Griba}
New experimental approaches, like high-resolution crossed molecular beam
experiments using Zeeman deceleration in combination with Velocity Map Imaging,
\cite{Plomp2021}  might provide detailed information about the title reaction
in the near future.
The paper is structured as follows. In the
next section, we will briefly describe the theoretical methodology. We also
provide details on the considered PES and the calculation of effective
potentials. The results from the dynamical calculations and the models to
understand them will be shown and discussed in section III, that includes the
state-to-state integral and differential cross section.  Finally, a summary of
the work and the conclusions will be given in Section IV.

\section{Theoretical Methods} \label{Themet}

\subsection{Potential Energy Surface}
%
\begin{figure}[b!]
\vspace{-0.5cm}
\includegraphics[width=60ex]{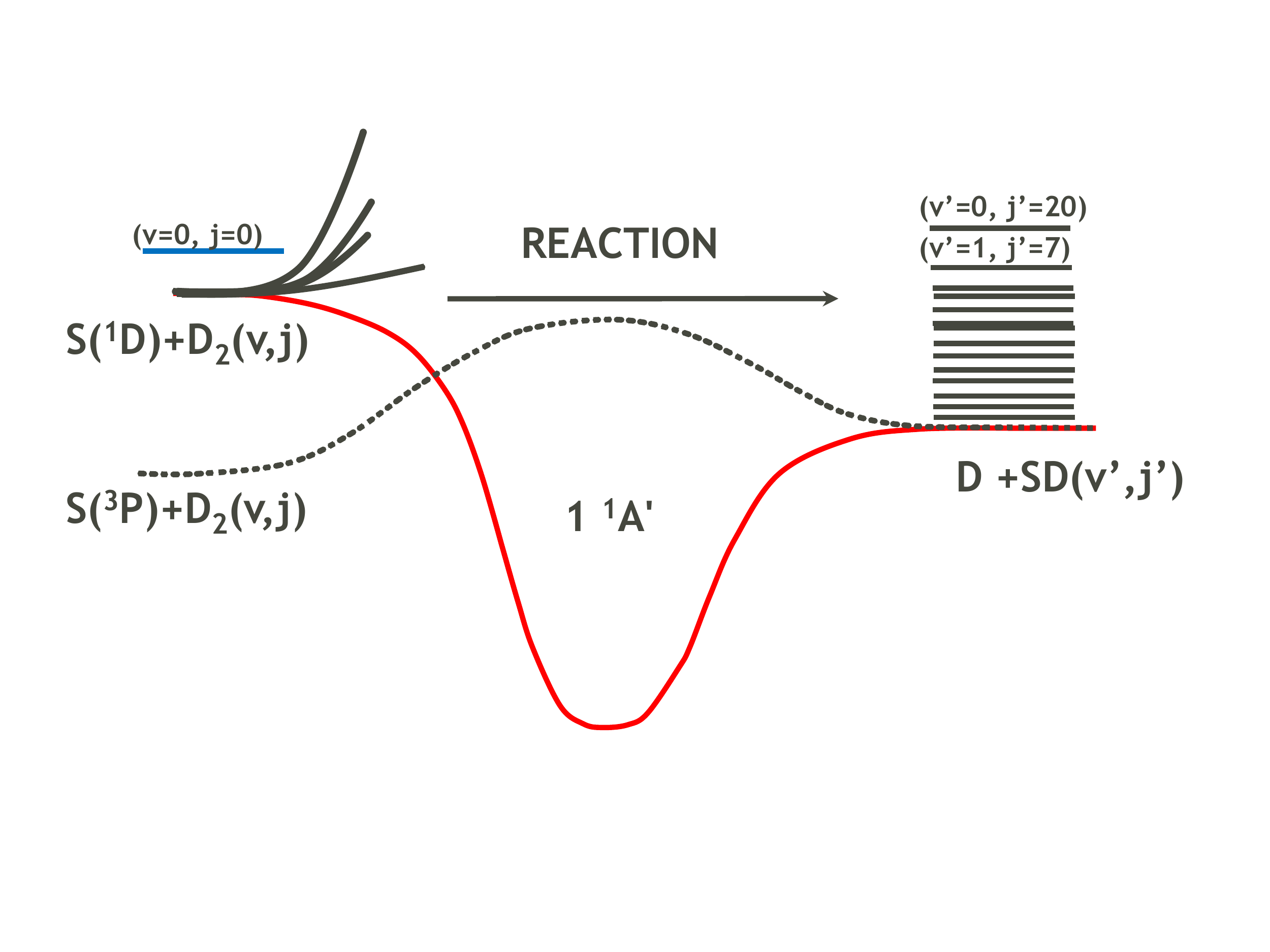}
\vspace{-1.5cm}
\caption{Potential energy surface sketch. At the total energy corresponding to the
$D_2(v=0,j=0)$ initial state, the $v'$=0 and $v'$=1 manifolds comprise 20 and 6 rotational states, respectively.}
 \label{fig1}
\end{figure}
Five singlet PES's correlate with the reactants  asymptote in the title reaction,
S($^{1}D$)+D$_2$. However, only two of them,  1$^{1}A'$ and 1$^{1}A''$  correlate
adiabatically with the ground state products, SD($^2\Pi$)+D. The 1$^{1}A''$ surface,
which corresponds to a direct abstraction process, has a barrier of 0.43 eV and can be
disregarded at low collision energy.  The ground state surface, the 1$^{1}A'$, features a
deep well in the reaction path ($\approx$4 eV depth),  and a relatively small
exoergicity ($\Delta_r H_0^o=-0.254$ eV). On this PES, the reaction takes place by
insertion of the sulphur atom through the D-D bond, leading to a long-lived
intermediate collision complex.  Although it has been suggested that the S($^{1}D$)
$\rightarrow$ S($^{3}P$) quenching  may play a significant role, quantum scattering
calculations  considering only  the $1A'$ PES can reproduce experimental data even at
low energies, \cite{Lara2,PhysRevLett.109.133201,LaraJPC2016} and this will be our
approach here. In Fig.~\ref{fig1}, we show a sketch  of the different PES's which
correlate with the considered state of the reactants. Apart from the deep well which
characterizes the 1$A'$ PES, let us note that there are many product rovibrational
states which are open at the energies explored in this work. In particular, at the total energies
corresponding to the cold and ultracold processes ($E_c \le 1$\,K) there are 20
rotational states with $v'$=0 and 6 rotational states with $v'$=1 accessible from the
complex.

In this and our previous works on this system at low kinetic energies, we have
used accurate LR interactions \cite{Lara2}, and merged them with  the
short-range interactions provided by the widely used {\em ab initio} RKHS PES
by Ho {\em et al.}~\cite{Skodje02}. The main terms which characterize the LR
interactions are the quadrupole-quadrupole  electrostatic term, varying as
$\sim R^{-5}$, and the dipole-induced dipole-induced dispersion one, $\sim
R^{-6}$. Indeed, the LR potential matrix elements, expressed in a diabatic
basis of five asymptotically degenerate atomic states, $|L \lambda \rangle $,
corresponding to the 5 values of the $\lambda$ projection of the sulphur
orbital angular momentum, $L$=2, on the body-fixed $z$-axis, can be written as:
\cite{C0CP02705G}
\begin{align}
\label{LR}
V_{\lambda,\lambda'}(R,\theta) =
&\frac{1}{R^5}  C_5^{\lambda,\lambda'}\;{\cal C}_{2,\lambda'-\lambda}(\gamma,0)
\nonumber \\
&
-\frac{1}{R^6}\sum_{k=0,2}C_{6,k}^{\lambda,\lambda'}\;{\cal
C}_{k,\lambda'-\lambda}(\gamma,0)
\end{align}
where $\gamma$ is the Jacobi angle, the angular functions ${\cal C}_{l,m}(\theta,\phi)$ are the modified
spherical harmonics, and the coefficients $C_5^{\lambda\lambda'}$ and
$C_{6,k}^{\lambda\lambda'}$ correspond to the electrostatic
(quadrupole-quadrupole) and dispersion (dipole-induced dipole-induced)
contributions, respectively.

As already mentioned, our treatment of the dynamics for the collision with D$_2$($j=0$) will
be adiabatic. However, if we considered a non-adiabatic treatment using 5 electro-nuclear
basis functions (each of them corresponding to an atomic $| L \lambda \rangle $ state),
which depend on the coordinates of both atoms and electrons,  the contribution of
quadrupole-quadrupole and dispersion anisotropic terms would be zero since the potential coupling
elements of the type $\langle j=0 | P_2(\cos \gamma) | j=0 \rangle$ vanish. Accordingly,  in
order to introduce a right dependence of the LR interaction with distance, we have matched
the PES by Ho, with a pure isotropic dispersion term of the type $-C_6 / R^{6}$ around a
distance of $R$=13.5\,a$_0$. The value of $C_{6,k=0}^{0 , 0}$, the lowest isotropic one, was
chosen to that purpose, $C_6$=41.78\,a.u.

\subsection{The effective potentials}
\label{effec}
As explained in Ref.~\citenum{Lara2} and \citenum{LaraJCP2015}, a convenient basis in order to
expand the nuclear wavefunction in the LR region can be  characterized by quantum
numbers $(J,M,v,j,l)$, with ($J,M$) the total angular momentum and its projection on
the Space-Fixed (SF) $Z$ axis, ($v,j$) the rovibrational quantum numbers of the diatom
and $l$ the relative orbital angular momentum. We represent  it as $\varphi^{J M}_{v j
l }$. Now, the matrix element of the electronic potential evaluated for the incoming
channel $(J,M,v_0, j_0, l_0)$, plus the centrifugal potential term, can be understood at
each distance $R$ as the effective 1D-potential, $V_{\rm eff}(R)$, felt by the colliding
partners while approaching:
\begin{equation}
V_{\rm eff}(R)=\langle \varphi^{J M}_{v_0 j_0 l_0} | V(R,r,\gamma)  | \varphi^{J
M}_{v_0 j_0 l_0} \rangle + \frac{l_0(l_0+1) \hbar^2}{2 \mu R^2}
\end{equation}
where $\langle ... \rangle$ indicates here the integration over the Jacobi
coordinates $r$ and $\gamma$ and $\mu$ is the reduced mass.  The evaluation of the potential matrix element, in brief
$V_{l_0,l_0}(R)$, involves the change to the {\em helicity basis}
(labeled by the projection $\Omega_j$ of ${\bm J}$ on the Body--Fixed (BF)
coordinate system, instead of $l$) using $3j$ symbols and leads to the following
expression:
\begin{align}\label{camio}
V_{l_0,l_0}(R)=&
2 \pi (2l_0+1) \,  \sum_{\Omega_j} {\left(
\begin{array}{ccc} j_0 & l_0 & J  \\
\Omega_j & 0  & -\Omega_j
\end{array}\right) }^2  \\
&\int  \chi_{v_0,j_0}^2(r) Y_{j_0 \Omega_j}^2(\gamma,0) V(R, r,\gamma) \; dr\;
d(\cos\gamma)
. \nonumber
\end{align}
\begin{figure}[b!]
 \begin{center}
\includegraphics[width=60ex]{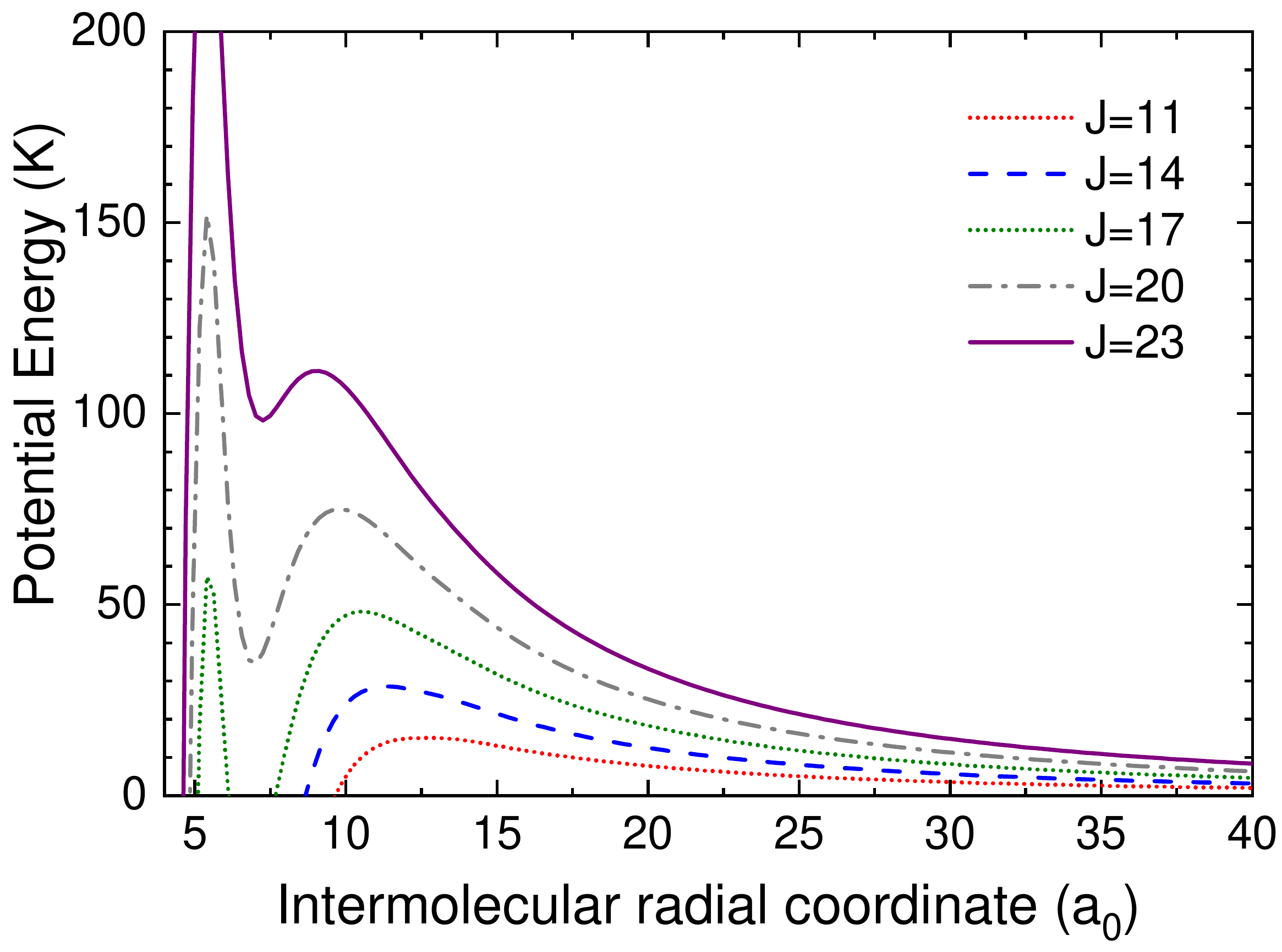}
\caption{Effective potentials (centrifugal plus electronic
  potential) for different values of the orbital (here also total) angular momentum.  Note the
  double-maximum structure.}
 \label{fig2}
\end{center}
\end{figure}
Effective potential for several $J's$ are shown in Fig.~\ref{fig2}. They exhibit a
double-maximum structure, which was already noted in previous works on the system
for other isotopic variants. \cite{lara:11,PhysRevLett.109.133201} While the outer
maximum is the well known centrifugal barrier, resulting from the increasing attraction
to the internal region of the PES and the repulsion of the centrifugal term, the inner
maximum has a dynamical character. The latter is due to the angular averaging in
$\gamma$ angle (at each $R$), the potential energy being increasingly negative closer
to perpendicular angles (insertion) and positive (with a potential barrier appearing) for
linear approach.\cite{LaraJCP2011} Let us remark that the double maximum structure
also appears in the effective potentials calculated using more recent PESs.
\cite{PhysRevLett.109.133201}  We have related this structure to the appearance of
shape resonances for high partial waves, which would be supported by the potential
well between both maxima. \cite{lara:11} The change of slope observed on previously
published experimental and theoretical excitation functions can be also attributed to
this double maximum structure.\cite{PhysRevLett.109.133201,LaraJPC2016}

\subsection{Dynamical methodology and calculation details}

The method based on hyperspherical coordinates  developed by Launay {\em et
al.}~\cite{Launayfirst} has been widely used for thermal and hyperthermal
reactive scattering \cite{Banares02,Honva,Banaresulti,hon04}. It uses
Smith-Witten hyperspherical democratic coordinates to describe the interaction
region of the configuration space and considers a single adiabatic PES. The
hyper-radius coordinate, $\rho$, is split into a multitude of sectors, and a
coupled channel method of the type diabatic-by-sector is used. The chosen basis
of surface functions (which depend on the hyperangles and are defined at the
centre of each sector) is particularly appropriate to study insertion
reactions with very deep wells. The log-derivative matrix built by expanding
the total wavefunction using the sector basis, is propagated outwards in
$\rho$, and changed of basis at the border of each sector. Its value at a long
enough value of hyper-radius, $\rho_0$, is matched to a set of suitable radial
functions, called asymptotic functions (AFs). At thermal energies, the AF s are
the familiar regular and irregular radial Bessel functions (accounting for the
presence of the centrifugal potential). To study cold and ultracold collisions,
the AF's were modified to account for both the centrifugal potential and the
particular LR potential existing at $\rho_0$~\cite{Lara2,hon04,Sol02,Quem04}:
in this way the AF's are adapted to each specific LR behavior, ensuring the
collisional boundary conditions while working at finite distance. This avoids
propagations in hyperspherical coordinates up to extremely large intermolecular
separations. We use the coupled-equation version of  the method of De
Vogelaere~\cite{Lester} to calculate the AF's in Jacobi
coordinates~\cite{Lara2} by solving a system of very few differential equations
(only one for the case $j$=0). Although propagations from very large radial
distances, (where LR interactions are negligible) up to $\rho_0$ are required
for their calculation, the expense is minimal in comparison with the
alternative of propagating directly in hyperspherical coordinates. This
implementation is applied here to deal with a system with $\sim R^{-6}$ LR
behavior. Let us note that to converge the elastic cross sections at a
collision energy of $10^{-8}$\,K, De Vogelaere's propagations starting from
separations of $\sim$10$^4$ a$_0$ were needed to calculate the AF's. However,
the computational cost of this calculation is affordable. Reaching so large
intermolecular separations using the unmodified (direct) implementation in
hyperspherical coordinates would be unthinkable. The AF's allow to 'bring' the
asymptote to reduced finite distances.

We have calculated quantum reactive cross-sections for the collision
S($^{1}D$)+ D$_2$($v$=0, $j$=0)$\rightarrow$ SD + D in the collision energy
range 10$^{-8}$\,K--220\,K. This range covers part of the thermal or
Langevin regime, where an appreciable number of partial waves are open and the
classical Langevin capture model might work (usually 4-5 partial
waves are enough), and the cold ($E<$1 K $\sim$ 0.7 cm$^{-1}$) and ultracold
($T<$1 mK) regimes. At the energies considered in this work, the Langevin
regime can be taken to be comprise the 2\,K-220\,K range, with 5 and 27
partial waves open, respectively. This can be concluded by examining the height
of the effective potentials. Partial
waves $J=$0-27 were needed in order
to converge the reactive cross sections. The elastic counterparts require
higher total angular momenta and they are only converged in partial waves up to
approximately 1\,K. The number of adiabatic channels in hyperspherical
coordinates included in the calculations lies in the range  from 380 ($J$=0) to
6605 ($J$=27). The propagation in hyperspherical coordinates was taken from
$\rho$=1.8 a$_0$ up to $\rho_0$=16.6 a$_0$, where the matching to the AFs was
performed. The integration to calculate the AF corresponding to the incoming
channel at each considered energy was taken from a radial distance where the
absolute value of the potential is $10^{-8}\times E$ up to $\rho_0$. This amounts to starting at
$\approx 10^4$ a$_0$ for the lowest energy considered. Such a stringent
criterion was necessary in order to obtain the correct Wigner behavior of the
elastic cross sections.

Our calculations yield the $S$-matrix as a function of the energy for each
value of $J$ or equivalently $l$, the relative orbital angular momentum (note
that in our case $j$=0). The complex (energy dependent) scattering length,
$\tilde{a}_{l}(k)=\alpha_{l}(k)-i \beta_{l}(k)$, essential ingredient when
considering the ultracold regime, can be evaluated using the elastic elements of
the $S$-matrix~\cite{Hutson07} by applying the expression:
\begin{equation}
\tilde{a}_{lm}(k)=\frac{1}{ik} \,\frac{1-S_{lm,lm}(k)}{1+S_{lm,lm}(k)}
\end{equation}
where $k$ is the wavenumber. At low enough collision energy, the complex scattering length relates to the elastic,
$\sigma_{\rm el}$, and total-loss cross-section, $\sigma_{\rm loss}$ (which means inelastic
$\sigma_{\rm in}$ plus reactive $\sigma_{\rm r}$) in the following way:
\begin{eqnarray}\label{losess0}
\sigma^{lm}_{\rm el} &\to& 4 \pi (\alpha_{lm}^2 + \beta_{lm}^2) \\
\sigma^{lm}_{\rm loss} &\to&  \frac{4\pi \beta_{lm}}{ k}
\label{losess}
\end{eqnarray}
As long as one is not interested in product-state resolved magnitudes, it must
be stressed that all the significant information is contained in the diagonal
elements $S_{l m,l m}(k)$ of the S-matrix:
\begin{eqnarray}
\sigma_{\rm el}&=&\sum_{l,m}\sigma^{l m}_{\rm el} =
\frac{\pi}{k^2}\sum_{l,m}|1-S_{l m,l m}(k)|^2 \\
\sigma_{\rm r}&=&\sum_{l,m}\sigma^{l m}_{\rm r} =
\frac{\pi}{k^2}\sum_{l, m}[1-|S_{l m, l m}(k)|^2]
\end{eqnarray}

\subsection{Classical Langevin Model}

The classical (Langevin) capture model for an $R^{-6}$ potential is used to rationalize the
cross sections and rate constants for  atom + molecule collisions at thermal energies.
\footnote{More precisely, this model should be named as `Gorin model' for the
particular case $n$=6; however, the generic name of 'Langevin model' is nowadays
used for any spherically symmetric $R^{-n}$ potential, and not only for the case $n$=4}
This model is approximately valid for collisions dominated by attractive potentials and
large reaction probabilities, usually providing an upper-limit to the reaction probability.
The assumption is that the reaction occurs whenever the reactants are able to reach
the inner part of the PES and interact strongly, that is, whenever the kinetic energy
overcomes the centrifugal barrier, and the reactants can get close enough. Besides, it
assumes a classical non-quantized orbital angular momentum.

The Langevin expressions for the reaction cross section, $\sigma_{\rm L}(E)$,
and rate coefficient, $k_{\rm L}(E)$, can be obtained starting from the quantum
expression
\begin{equation}
\sigma_{\rm r}(E)= \sum_{l=0}^{\infty} \sigma^{l}_{\rm r}(E)= (\pi /
k^2) \sum_{l=0}^{\infty} (2l+1) P_{\rm r}^{l}(E),
\end{equation}
where $P_r^{l}(E)$ is the reaction
probability for a particular partial wave $l$ and $m$ projection
(we will  use $P_r^{l}(E)$ or
$P_r^{J}(E)$ indistinctly in the notation below ($j=0$). By: (i) calculating
the centrifugal barrier height for each $l$ corresponding to an analytical
potential $-C_6/R^6$; (ii) assuming that $P_r^{l}(E)=1$ if the kinetic energy
is higher than the centrifugal barrier, and null otherwise; and (iii) replacing
the summation with an integral over continuous values of $l$ (valid for high
enough maximum $l$) we obtain the Langevin cross section,
\begin{equation}\label{Langevincross}
\sigma_{\rm L}(E)=3\pi(C_6 /4E)^{1/3},
\end{equation}
and, multiplying by the relative velocity, the Langevin rate coefficient
\begin{equation}\label{Langevinrate}
k_{\rm L}(E)= 3 \pi (C_6^2 E /2 \mu^3)^{1/6}.
\end{equation}
These expressions can only be valid when many partial waves are open. In fact, in
the limit of vanishing kinetic energy (only $l=0$ is involved) the right Wigner
behavior, $\sigma_{\rm r}(E)\sim E^{-1/2}$, is not recovered.

The accuracy of the Langevin  model can be improved in reactions mediated by a
complex by:  (i) calculating the heights of the centrifugal barriers, using the
effective potentials, thus accounting for deviations from the $R^{-6}$ behavior
at short-range; (ii) carrying out the summation over discrete partial waves
instead of an integral over a continuous variable; and (iii) correcting the
assumption $P_{\rm r}^{l}(E)=1$ with a statistical factor, $P_{{\rm complex}\to
{\rm prod}}$, that is calculated as the ratio between the number of product
states and the total number of states (reagents and products), and accounts for
the fact that the probability of a complex to decompose into the products
arrangement channel will be usually lower than 1. We named this improved model
{\em Numerical-Capture Statistical} (NCS) model in previous
works~\cite{LaraJCP2015}.

In contrast to the D$^+$+ H$_2$ reaction, where $P_{{\rm complex}\to
{\rm prod}}$ was significantly lower than 1~\cite{LaraPRA}, there are many SD
channels accessible from the well for the  S+D$_2$ reaction (see Fig.~\ref{fig1}),
and hence the probability flux is
expected to be irreversibly lost from the incoming channel due to the numerous
couplings inside the well which lead to the product channels.\cite{LaraJCP2015,Lara2}
Consequently, most of the complexes will decompose to form the products and very
few to give back the reactants. In other words, the NCS model assumes $P_r^{J}(E)=1$
in this system, as the Langevin model does, and both models should only differ in the
discrete distribution of impact parameters considered by the former.

As the Langevin model, the NCS model does not provide the right behavior in the
limit of vanishing kinetic energy. When only s-wave contributes at low enough kinetic
energy, and assuming that  $P_{\rm r}^{l=0}(E)=1$, we get:
\begin{equation}
\sigma_{\rm r}^{l=0}(E)=  \frac{\pi}{k^2} P_{\rm r}^{l=0}(E) =\frac{\pi}{k^2} \propto E^{-1}
\end{equation}
The wrong power stems from the assumption $P_{\rm r}^{l=0}(E)=1$, which clearly
does not fulfill the limiting behavior of the reaction probability, that should be $P_{\rm
r}^{l=0}(E)\propto E^{1/2}$ as $E\to 0$. Due to the LR forces the incoming flux is
reflected back before reaching the SR region of the PES even in the absence of
centrifugal barrier, and hence the reaction probability goes to zero. It must
be, however, stated that the Langevin's assumption: ``all the flux that reaches the
transition state leads to reaction" is not wrong. What is wrong is the mathematical
implementation of this assumption in the classical model: $P_{\rm r}^{l=0}(E)=1$ at any energy.
There exist other ways to implement Langevin's idea without violating the Wigner
threshold laws using the MQDT.

\begin{figure}[h!]
 \begin{center}
\includegraphics[width=65ex]{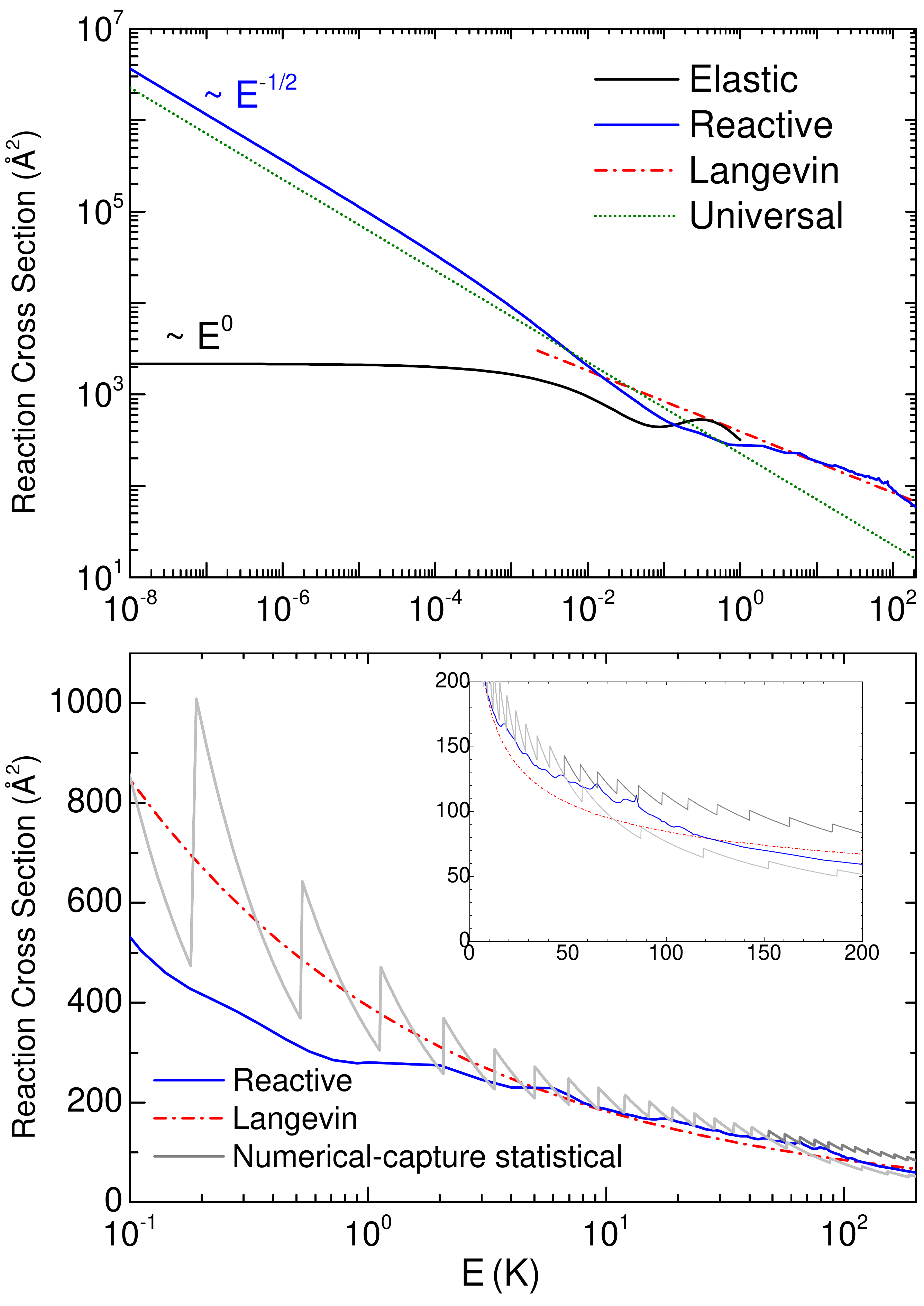}
\caption{The upper panel displays in logarithmic scale the QM total reaction and elastic cross-sections, whose
Wigner limits are $E^{-1/2}$ and $E^0$, respectively. The reaction cross-section
is compared with the universal prediction at ultra-low collision energies
and with the Langevin prediction (with the right $C_6$), when many partial waves are open.
The lower panel shows the reaction cross sections at the Langevin-regime collision energies.
The QM cross sections are compared with the ones obtained using the classical Langevin
formula, Eq.~\eqref{Langevincross}, and the more realistic numerical-capture
statistical model. The inset highlights these latter results in a linear scale of energies. }
\vspace{-1cm}
\label{fig3}
\end{center}
\end{figure}

\subsection{Calculating capture probabilities}
Under the influence of a particular potential, i.e. $V_{\rm LR}(R)=-C_6/R^{-6}$, within a region
where the WKB aproximation is valid, one can distinguish between incoming and outgoing
probability terms in the continuum wavefunction:
\begin{equation}
  \psi(R) \sim   \frac{ A \exp[- i  \int_{R} k(x) dx]  +
  B \exp[+ i \int_{R}  k(x) dx]}{\sqrt{k(R)}} \,,
\end{equation}
where $k(R)=[2\mu (E-V_{\rm LR}(R)]/\hbar$ is the wavenumber at a distance $R$ and $E$ is
the collision energy. The formula implies that part of the incident flux  is reflected by the
potential and part of it is transmitted. However, one can mathematically impose the absence of reflected term
(semi-classical condition for perfect absorption or transparency) at a
particular point in space, $R_m$: the logarithmic derivative of the wavefunction should take
there the simple form $Z(R_m)=-ik$. We say that the flux is captured by the potential
and we know this probability as {\em capture probability}. In order to determine it, we need
to find the solution of the 1D time-independent Schr{\"o}dinger equation (TISE), with
potential $V_{\rm LR}(R)=-C_6/R^{-6}$, which fulfills the condition $Z(R)=-ik$ at a short
matching distance. Once obtained, the S-matrix element associated to this solution will not
have modulus 1, due to the loss of flux at short distances. The lack of unitarity, $1-|S|^2$,
will be precisely the capture probability we want to calculate. In this work we use the method
by De Vogelaere \cite{Lester} to solve the TISE. More details about the procedure are given in
the Appendix.

\section{Results and Discussion}

\subsection{The reaction in the Langevin regime}

The QM reaction cross section calculated in the low energy region is shown in the bottom
panel of Fig.~\ref{fig3} along with the Langevin and NCS results. Interestingly, the Langevin
model does not always overestimate the QM result in the range 2-200\,K, where enough
partial waves are open for the integral to be a good aproximation to the discrete summation
in partial waves. Moreover, the slope of the Langevin model, which reveals the pace of the
opening of partial waves corresponding to a $R^{-6}$ potential, is not in agreement with the
slope of the QM cross section below 120\,K (see inset of Fig.~\ref{fig3}). Indeed, the
$C_6/R^6$ potential used in the Langevin model is not valid at short radial distances where
the centrifugal barriers for moderate $J$ values are located.
\begin{figure}
 \begin{center}
\includegraphics[width=65ex]{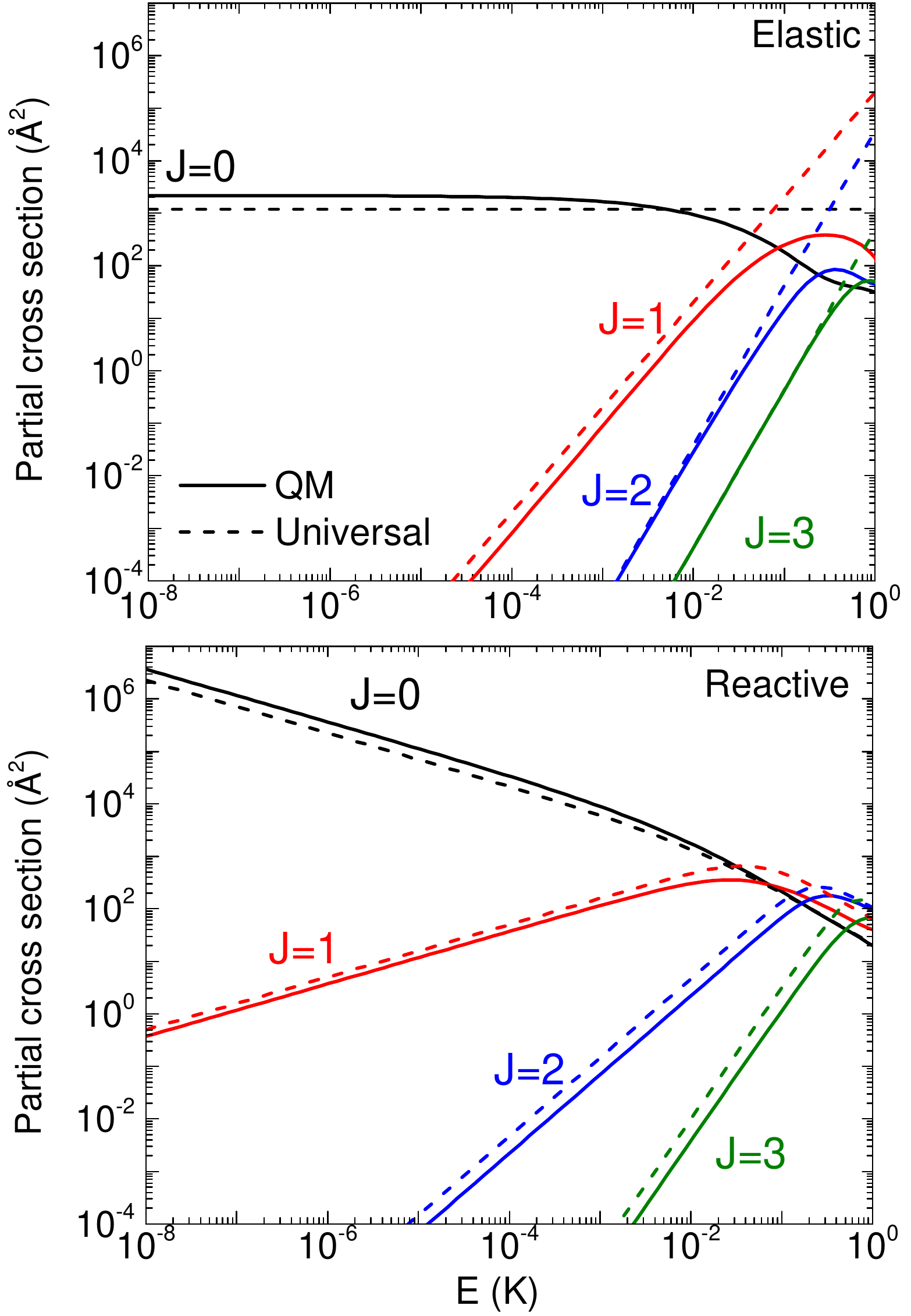}
\caption{Elastic (upper panel) and reaction (lower panel) cross-sections corresponding to the
 lowest partial waves for  S($^{1}D$)+ D$_2$ ($v$=0, $j$=0) collisions in the cold and
 ultracold regimes. The  QM results are compared with the results of the {\em universal}
 predictions (dashed line).}
 \label{fig4}
\end{center}
\end{figure}
The NCS model corrects the Langevin model by using the right effective potentials given
by Eq.~\eqref{camio} and depicted in Fig.~\ref{fig2}.  Accordingly, it should provide both an
upper limit and the right dependence of the cross-section with the collision energy. Due
to the double maximum structure of the effective potentials, special care should be taken when implementing
the NCS model; strictly speaking, the highest maximum should be chosen as the one
limiting the access of the probability flux to the interaction region. While the outer
maximum is higher for lower partial waves, $J<$ 17, it is the inner (dynamical) barrier
the one which allegedly limits the reactivity for $J \ge$ 17. The result of this
implementation is shown in solid grey line in Fig.~\ref{fig3}. The agreement with the
QM results is much better than the one obtained using the Langevin expression, both
working as a higher bound and following the slope of the QM results up to energies
$\sim 50$K (precisely the height of the two barriers associated to $J$=17). At these
energies, right where the inner maximum plays a role, the NCS model shows a
significant decrease and a sudden change of the average slope, a behavior which is
observed in the QM results above $E \approx 90$K. It seems that the effect of the inner
maximum is not reflected yet in the QM results for the partial waves in this energy region.
To understand this effect,  we calculated the NCS model cross sections by considering only
the height of the outer maxima, irrespective of whether it is the highest of the two.
The results obtained are shown using green colour line in the lower panel of
Fig.~\ref{fig3}, and are better appreciated in the inset. This second NCS implementation
is closer to the QM results up to energies around 90\,K, although it is unable to follow
the change observed in the QM results at higher energies. Interestingly, the QM value
seems to lie between the results of both implementations of the model. This is not
difficult to understand in view of the effective potentials: even if the inner barrier is
higher than the outer one for $J \ge$ 17, it is very narrow and the tunneling through it
is probably very significant. In this scenario, the flux seems to  be effectively limited by
the outer maximum up to $J \ge$ 21. For higher $J$'s, neither the outer nor the inner
barrier heights seem able to characterize the opening of partial waves: the effect of the
inner barrier is too significant to be ignored, but the tunneling is also too high to imply
that the opening of a partial wave occurs necessarily above it. It is the interplay of both
barriers which leads the process.

\subsection{Low partial waves in the (ultra)cold regime}

To analyze the behavior in the (ultra)cold regime for a LR potential of the form
$-C_n/R^n$ $(n > 3)$, it is useful to define a characteristic length, $R_n=(2 \mu C_n
/\hbar^2)^{1/(n-2)}$, and a characteristic energy, $E_n=\hbar^2 /(2 \mu R_n^{2})$. For
the case $n$=6 we get $R_6=(2 \mu C_6 /\hbar^2)^{1/4}$ and $E_6=\hbar^2 /(2 \mu
R_6^{2})$, which amount to $R_6=27.2\,a_0$ and $E_6/k_BT \approx
3.3\times10^{-2}$\,K, respectively, when evaluated for the title system \footnote{The
characteristic energy is of the order of the height of $p$-wave centrifugal barrier, and
appears around $R_n$. Indeed, the accurate position and height of the centrifugal
barrier in the case $n$=6 are $R_{\max}=(3 \mu C_6/\hbar^2)^{1/4} \approx 1.1 R_6$
and $E_{\max}=(2/3^{3/2})(\hbar^6 / \mu^3 C_6)^{1/2} \approx 1.1 E_6$,
respectively}. The so-called mean scattering length $\bar{a}$ \cite{Griba}, useful to
rationalize the problem, is given in this case by
\begin{equation}
\displaystyle \bar{a}=\frac{2 \pi}{\Gamma(1/4)^2 } R_6 \approx 13\,a_0 \,. \label{mean_a}
\end{equation}

In the wilderness of the ultracold regime, the well known Wigner threshold
laws~\cite{Sade,Weiner} work as a compass. They state that the elastic,
$\sigma_{\rm el}^{l}$, and the total-loss (inelastic plus reactive) cross
section, $\sigma_{\rm loss}^{l}$, associated to each partial wave vary, close
to threshold, as:
\begin{eqnarray}\label{normal}
\sigma_{\rm el}^{l} &\sim&  E^{2l} \\
\sigma_{\rm loss}^{l} &\sim&  E^{l-1/2}
\label{Wigner}
\end{eqnarray}
Given that H$_2$($v$=0, $j$=0) is the only rovibrational state of the reactants open at
the considered energies, the inelastic process is absent and losses are only associated to
reaction, $\sigma_{\rm loss}^{l}=\sigma_{\rm r}^{l}$. Besides, the threshold laws for
elastic scattering are modified for a potential with $n=6$. The phase shift for $l > 1$ at
very low collision energies is dominated by a term $\sim E^2$ originating from the
dispersion potential.\, \cite{Sade}  The anomalous behavior of the elastic cross section
is given by
\begin{eqnarray}
\sigma_{\rm el}^{0} &\sim&  E^{2l} ~~\text{for}~ l=0,1 \\
\sigma_{\rm el}^{l} &\sim&  E^{3} ~~\text{for}~ l>1
\end{eqnarray}
In summary, while the partial reaction cross sections are expected to change as
$E^{l-1/2}$, the partial elastic ones will remain constant for $l=0$, change
linearly with energy for $l =1$ and behave as $E^3$ for $l > 1$.

These predicted behaviors can be distinguished in Fig.~\ref{fig4}, where the reactive
(lower panel) and elastic cross sections (upper panel) for the lowest four partial waves,
are shown in solid line and log scale. Note the change of slope of the
$p$-wave ($J$=$l$=1) reaction cross section at energies around $E_6$ (33mK), close to the height of the
centrifugal barrier. It is also remarkable the anomalous behavior for $\sigma_{\rm
el}^{l}$, with the slope of the elastic cross sections being the same for $l = 2,3 $. A lack
of all the previous limit behaviors in the numerical results would have been taken as an
indication that a particular PES is not reliable to describe low energy collisions or that
the calculations are not well converged.

The s-wave contribution will dominate in the limit of extremely low kinetic
energies, where only $l=0$ is open. The total ({\em vs.} partial) reaction
cross section should change as $E^{-1/2}$ (the reaction rate coefficient being
thus constant) and the total elastic cross section remain constant (the elastic
rate coefficient changing as $E^{1/2}$). Note these behaviors in the upper
panel of Fig.~\ref{fig3}, where total reactive and elastic cross sections are
shown. The Langevin prediction is also shown for comparison.
Below the range of energies where the number of
open partial waves is noticeable, it largely separates from the numerical
results and it does not reach the right slope in the ultracold
regime~\footnote{Only for the case $n=4$ the energy dependence of $\sigma_{\rm
L}(E)$ meets the requirements of the Wigner threshold law, $\sigma_{r} (E) \sim
E^{-1/2}$, in the zero collision energy limit. Hence the formula has not even
sense in that regime for any other $n$}. Interestingly enough, this divergence
does not necessarily mean that the Langevin assumption (the reaction occurs
with almost unit probability when the reactants get close enough) is wrong, but
that the probability that the reactants get close enough to react is very low.
We will analyze this in the following section.

\subsection{The Quantum Langevin behavior}

We will consider the {\em quantal version} of the Langevin model proposed in the
context of MQDT in Ref. \,\citenum{Jach:02} (see Appendix for details).
The model, that complies with the Wigner laws,  allows us to move smoothly
from high energies to the ultracold regime under the same assumption: all the flux
reaching the SR region leads to reaction. This Langevin-type assumption,
translated into the language of the model, is expressed as: {\em the loss probability at
SR}, $P^{\rm re}$, is one, which is known as the universal behavior. Let us recall that
$P^{\rm re}$,  which is related to the $y$ parameter (see Appendix), denotes
the flux that is irreversibly lost from the incoming channel at SR, and is not directly
observable nor necessarily coincides with the reaction probability, $P_r^J(E)$. Expressed
in simple terms, the incoming flux which leads to reaction has to overcome
two obstacles: (i) it needs to reach the SR or transition state region without being
reflected backwards by the LR or centrifugal potential; (ii) it has to find its way from
the SR region to the products valley; this latter process is the one accounted for by the
$P^{\rm re}$ parameter.   We can conclude that only at high energies, when quantum
reflection by the LR potential and tunneling through the centrifugal barrier are
negligible, $P_r^J(E) \approx P^{\rm re}$. Furthermore, the expression
\begin{equation}
\sigma_{\rm r}(E) \approx P^{\rm re} \cdot \sigma_{\rm L}(E)
\label{sig2}
\end{equation}
would be valid in the Langevin regime if $P^{\rm re}$ were weakly dependent on the
energy and the partial wave~\cite{Jach:02}. According to Fig.~\ref{fig3}, the title
reaction fulfills Eq.~\ref{sig2} in the limited range 5\,K-15\,K with $P^{\rm re}=1$.
However, the deviation of $\sigma_{\rm L}(E)$ from $\sigma_{\rm r}(E)$ at higher
energies was not due to the failure of the Langevin's hypothesis, but to the inability of a
pure $R^{-6}$ term to account for the heights of the centrifugal barriers (which occur
at short distances) for higher $J$'s. In essence, it would be more precise to substitute
$\sigma_{\rm NCS}$ (with statistical factor 1) for $\sigma_{\rm L}$ in equation
Eq.\eqref{sig2}, hence finding that $P^{\rm re}$ is closer to 1 in a much wider range.
Besides, in a previous work we established a link between $P^{\rm re}$ and the
statistical factor $P_{{\rm complex}\to {\rm prod}}$, and hence a way to estimate the parameter
$y$.  \cite{LaraPRA}  As indicated above, we can assume that $P_{{\rm complex}\to {\rm prod}}\approx 1$
in the title system, leading again to $P^{\rm re}\approx 1$ and $y
\approx 1$. In conclusion: we can consider the system as roughly
{\em universal} or {\em quantum-Langevin}.

It would be interesting to compare the QM results with the behavior of a universal
system. Let us focus first in the reaction cross-sections, which in the case $y=1$
are given by the capture probabilities. Using De Vogelare's method and
proceeding as described in the Appendix, we have calculated the capture probabilities,
as a function of the energy, corresponding to a pure $-C_6/R^{6}$ potential for the
lowest partial waves; using them we have obtained the partial cross-sections. They are
shown in dashed line in the lower panel of Fig.\ref{fig4}, where they are compared with
the QM ones. The behavior for $J$=0 and $J$=1 in the ultracold limit is given by
Eq.~\eqref{unot} and Eq.~\eqref{trest}, stemming from Ref.~\citenum{Jach:03}.  The
limiting ratios  of the QM and capture values of the cross-sections as $E \to 0$ are 1.6,
0.7, 0.5, 0.4 and 0.8 for $J$=0, 1, 2, 3 and 4 respectively. Note that for $J>0$ the
cross-sections show a maximum as a function of the energy which is related to the
height of the centrifugal barrier. As only for low partial waves the centrifugal barrier is
well reproduced by a pure $-C_6/R^{6}$ asymptotic potential, the maxima in the QM
and the capture calculations show increasing differences as $J$ grows.

\begin{figure}[h!]
 \begin{center}
\includegraphics[width=65ex]{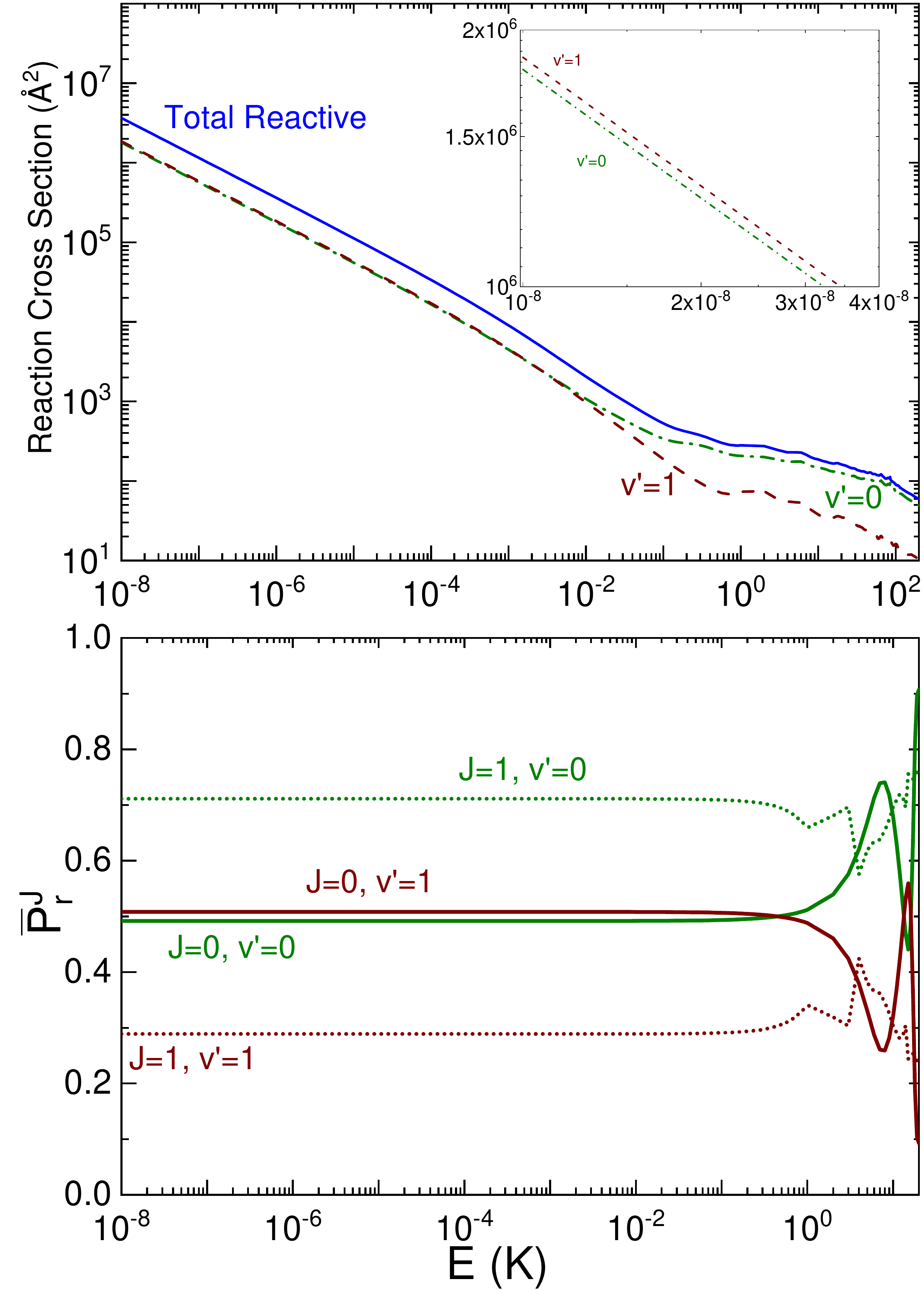}
\caption{Upper panel: Total reaction cross section as a function of the collision energy (blue
solid line) together with the
vibrationally resolved cross sections ($v'$=0 green and $v'$=1 brown dash-dotted lines).
For $E < 10^{-2}$\,K the cross section into $v'$=1 is larger than that
into $v'$=0 (see inset). Lower panel: fraction of collisions at a given $J$  leading to $v'$=1
(brown line) and to $v'$=0 (green line)
for total angular momentum $J$=0 (solid line) and $J$=1 (dashed line). At $E>5\cdot
10^{-1}$\,K the fraction into $v'$=0 takes over and even for $J$=0 no vibrational population inversion is found. Note that the
contribution of $J$=1 is negligible for
$E < 10^{-3}$\,K (see fig.\ref{fig4}).}
 \label{fig5}
\end{center}
\end{figure}

Regarding the elastic cross-sections, Ref.~\citenum{Jach:03} only provides their limit
behavior for $J$=0 and $J$=1, Eq.~\eqref{dost}
and Eq.~\eqref{cuatrot}. It is possible to generalize these expressions and calculate the
partial cross sections for any partial wave with anomalous behavior using
Eq.~\eqref{losess0}: $\alpha_{lm}$ can be obtained using the Born-approximation (see
Eq.~\eqref{elasalfa} the Appendix); $\beta_{lm}$ is directly related to the numerical
capture probabilities we have calculated using Eq.~\eqref{losess}. The universal partial
elastic cross-sections are shown in dashed line in the upper panel of Fig.\ref{fig4},
where they are compared with the QM ones. The limit ratios of the QM results and the
universal values are 1.8, 0.4, 0.9, 1 and 1 for $J$=0, 1, 2, 3 and 4 respectively.


The agreement between the QM results and the universal ones is on the same order of
magnitude of the one we obtained in our previous study on the system D$^{+}$+ H$_2$
when applying the MQDT model. What is remarkable is that the QM reaction
cross-section for $J$=0 is higher than the universal prediction, which implies that the reaction
probability is higher than the total capture. One of the interesting predictions of the
MQDT model is precisely that the reaction probability at ultracold energies is not
limited by the capture or probability transmitted by the LR potential. In fact, by
recalling that the reaction probability can be related to the imaginary part of
the scattering length, $P_r^l(E)=4 k \beta_l$, and using Eq.~\eqref{unap}, we get the
following limit behavior for $l=0$ in the general, non-necessarily universal, case:
\begin{equation}
 P_r^{l=0}(E)=4 k \,\beta_0(k) = 4 k \bar a \times y \, \frac{1+(1-s)^2}{1+y^2 (1-s)^2}
\label{capture}
 \end{equation}
The first factor, $4 k \bar a$, is the limit value of the capture probability ($P_r^{l=0}$
when $y$=1), and the second factor can be higher than 1 for values of $y \ne 1$, and
for high enough values of $s$, what can be usually associated to the presence of
resonant states close to threshold.  We conclude that the reaction
probability can be higher than the capture probability, however counterintuitive this
idea may appear. This seems to be the behavior in the title system (see the
corresponding cross-sections in the upper panel of Fig.~\ref{fig3}).

\begin{figure*}[ht!]
 \begin{center}
\includegraphics[width=80ex]{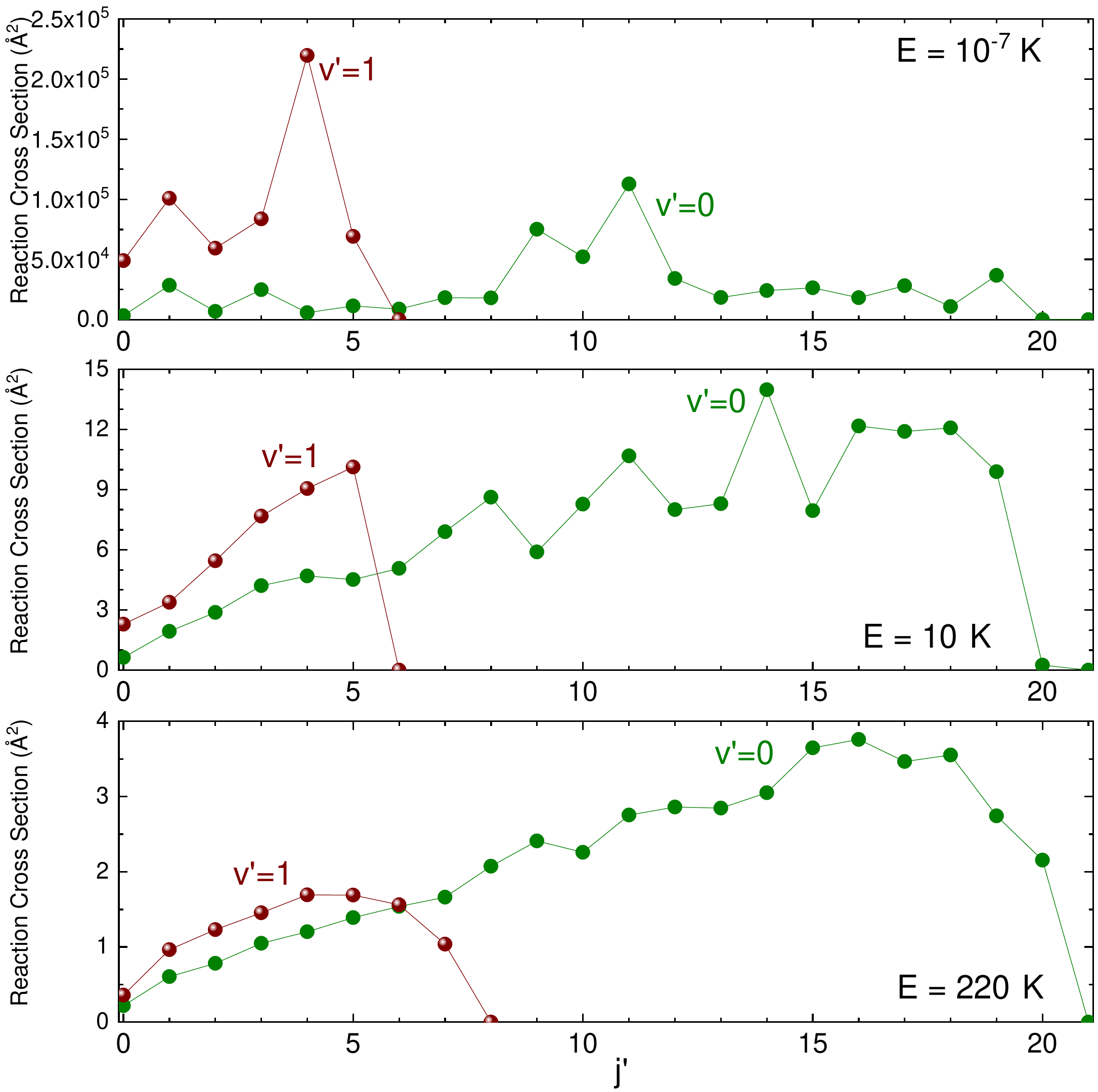}
\caption{State-to-state integral reaction cross-sections at 10$^{-7}$ \,K (upper panel),
10\,K (middle panel)
and 220\,K (bottom panel) collision energy for $v'$=0 (green) and $v'$=1. }
 \label{fig6}
\end{center}
\end{figure*}

\subsection{The rovibrational state distribution in the ultracold regime}

\begin{figure}[ht!]
 \begin{center}
\includegraphics[width=50ex]{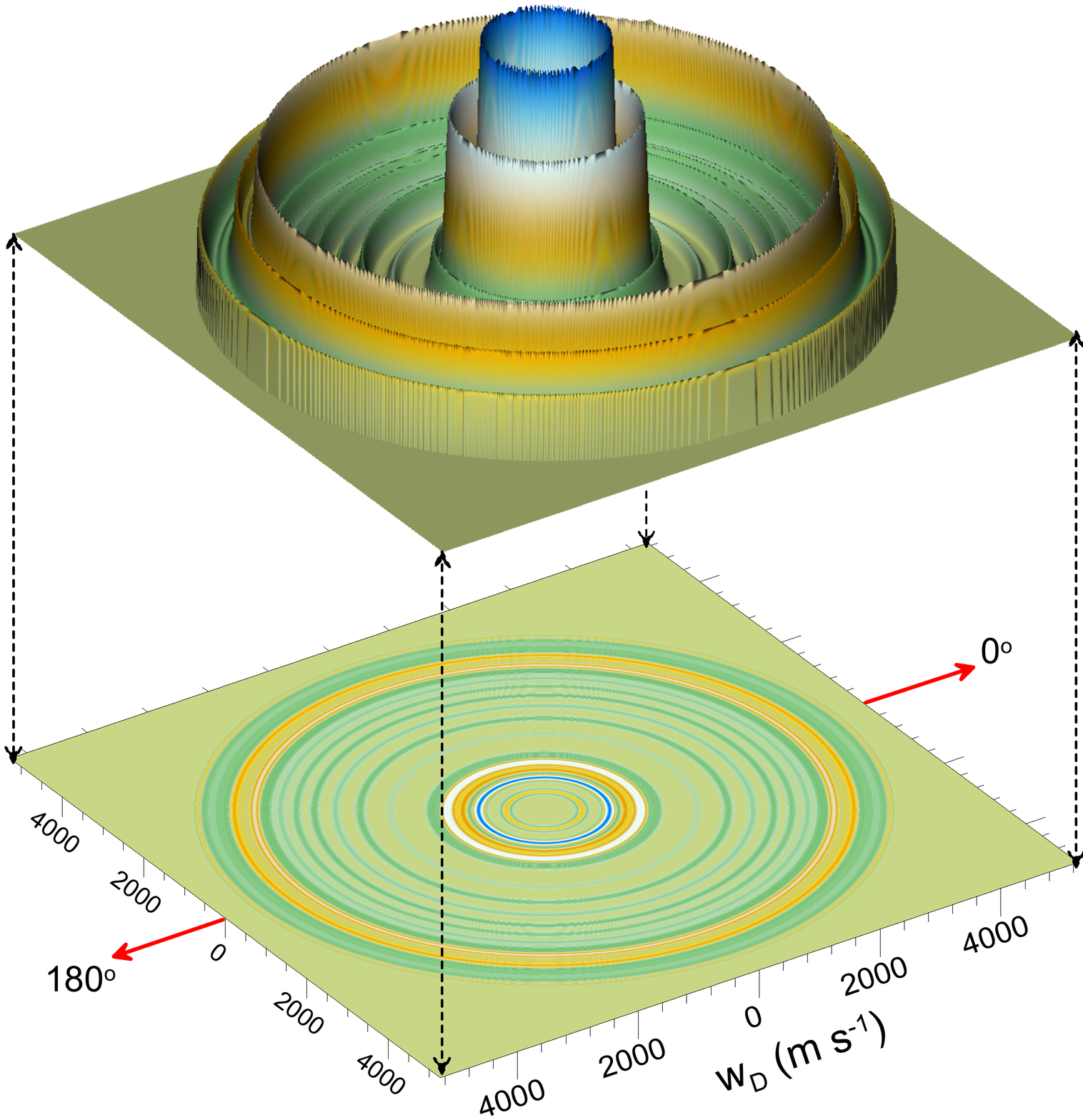}
\includegraphics[width=60ex]{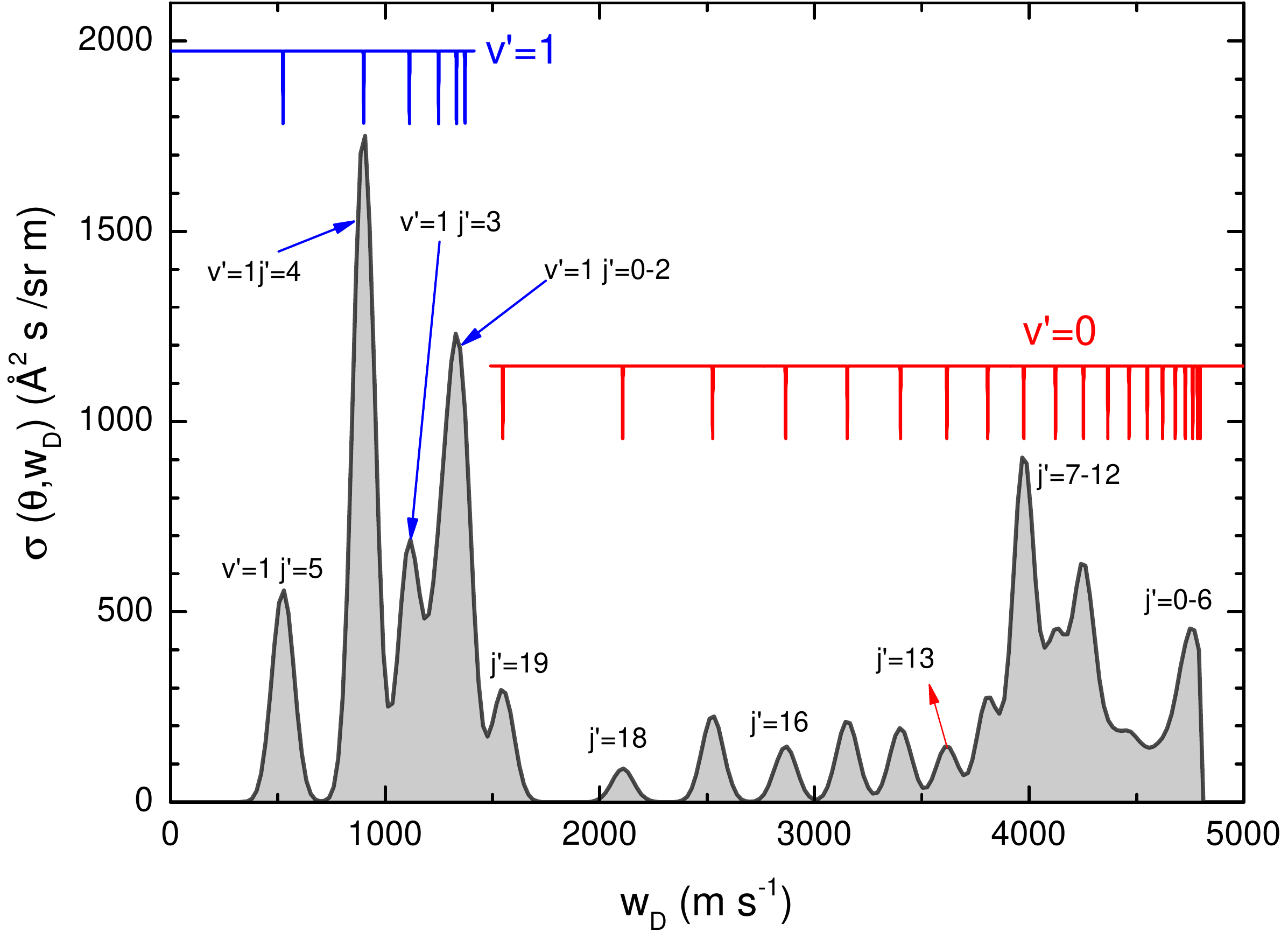}
\caption{Upper panel: Polar map showing the angular and recoil velocity distribution of the
scattered D product in the center-of-mass frame
at the collision energy of 10$^{-7}$\,K.  At this collision energy the only significant
contribution proceeds from $J$=0; hence the differential cross section is isotropic.
Lower panel: D-atom distribution of recoil
velocities corresponding to the indicated SD rovibrational states at any angle.
As can be seen, the intensity of the peaks correlating with the SD rotational states of $v'$=1
prevail over those of $v'$=0. The comb
shows the center of D-atom recoil that corresponds to the various internal states of the SD
co-product. An arbitrary width of 120 m s$^{-1}$ has been employed in the Gaussian functions that
mimic the D velocities associated to the various
SD  rovibrational states.}
\label{fig7}
\end{center}
\end{figure}
The availability of accurate QM results for the system provides a good opportunity to
investigate whether the final state distribution complies with the statistical assumption
in the ultracold regime.
Specifically, the S($^1D$)+H$_2$ reaction and its isotopic variants have been
considered as benchmarks of the  statistical model assumptions: the transition
probability from an initial $i$ state to a final $f$ state is expressed as the product of
the capture probability from channel $i$ and the fraction of collision complexes that
decay into the product channel $f$; that is,
\begin{equation}
P_{i \to f}=P_i^{\rm cap} \cdot P_{{\rm complex} \to f}
\end{equation}
The breakdown of the collision complex is essentially ergodic and hence the spacial and
state distributions are essentially statistical. Statistical models
\cite{RGM:JCP03,Guo:JCP2005,AGS:JCP08,G-L:IRPC2007,Guo:IRPC2012} applied to
S($^1D$)+H$_2$ have shown that at thermal and hyperthermal energies the overall
(coarse-grained) description of the reaction is in very good agreement with accurate
QM scattering calculations.\cite{Banares02,Banaresulti,LaraJCP2011,LaraJCP2012}
However, an important issue to be discerned is whether the statistical characterization
of the reaction holds at the cold and especially at the ultracold regimes.

The upper panel of Fig.~\ref{fig5} shows the vibrationally resolved reaction
cross-section that includes the contributions from all the $J$ partial waves. As can be
seen, the energy dependence of the vibrationally resolved cross sections follows the
Wigner law. In addition, quite remarkably, a vibrational population inversion is
observed at ultracold energies below $10^{-2}$\,K, where $J$=0 prevails (see inset), at
variance with the  expectations of the statistical model. At higher collision energies,
however, the ground vibrational state of SD becomes more populated than $v'$=1. To
further investigate this effect, the lower panel of Fig.~\ref{fig5} displays the relative
fraction of reactive collisions, ${\bar P}^J_r=\sigma(v')/\sum_{v'}\sigma(v')$, leading to
$v'$=0 and $v'$=1 for $J$=0 and $J$=1 as a function of the collision energy. For $J$=1
the relative probability is constant at collision energies below 0.2\,K,  and
$\approx$30\% of the collisions appear in $v'$=1. In contrast, for $J$=0 the fraction of
collisions into $v'$=1 is $\approx$52\%, being also constant at lower energies.  At energies
above 0.2\,K,  ${\bar P}^{J=0}_r$ for $v'$=1 decreases rapidly, and except for some
specific energies, the population inversion no longer takes place.

The next step is to investigate the rovibrational distributions, $\sigma_r(v',j'; E)$, of the
SD product at energies in the ultracold, cold and thermal regimes. Fig.~\ref{fig6}
displays the rotational distributions for $v'$=0 and $v'$=1 at  $10^{-7}$\,K  (upper
panel), 10\,K (middle panel) and 220\,K (bottom panel) collision energies. At
$10^{-7}$\,K $J$=0 is the only contribution. At 10\,K  $J_{\max}$ is 11. In both cases,
the highest rotational states that can be populated are $j'$=19 and $j'$=5 for $v'$=0
and $v'$=1, respectively, since the available energy proceeds from the exoergicity of
the reaction. At 220\,K $J_{\max}$=25, exceeding the value of the highest accessible
$j'$ state ($j'$=20)  for $v'$=0. Relevant to these three cases is the general preference
for populating $v'$=1 when considering the same $j'$ value. Indeed,
at the three energies shown in Fig.~\ref{fig6},
the cross sections for $j' \le 5$ are larger for $v'$=1. Moreover, at ultracold energies,
this preference is large enough that  when summed over rotational states, there is an
slight vibrational inversion for $v'$=1 products.

The main assumption of the statistical model is that all the states compatible
with parity, conservation of energy and angular momentum are equally
populated. A somewhat more refined version, includes the surmounting of the
effective barrier (including the centrifugal term). An approximate expression of the cross
section for the statistical model when the only initial state is $v$=0,
$j$=0 (hence even diatomic parity and $(-1)^J$ triatomic parity), can be written as:
\begin{align}
\sigma(v',j'; E)= &\frac{\pi}{k^2} \sum_J (2J+1)\times\nonumber\\
& p^J_{0,0} \frac{[\min(J,j')+1] \Theta[E'_T - V_{\rm eff}]}{D^J}\,,
\label{sm}
\end{align}
where $p^J_{0,0}$ is the capture from the D$_2$($v$=0,
$j$=0) initial state, $\Theta(E'_T -
V_{\rm eff})$ is the Heaviside function such that if the product's
translational energy, $E'_T(v',j')$,  is smaller than the effective barrier
in the exit channel,
whose centrifugal term depends on $l'$, no reaction takes place.  The
denominator $D^J$ is  the sum of capture probabilities to both the initial
state and all the product states:
\begin{align}
D^J=p^J_{0,0} + \sum_{{\rm} all \,\, v',j'} [\min(J,j')+1]\Theta[E'_T - V_{\rm eff}]
\end{align}
The crucial term is $[\min(J,j')+1]$, that represents the number of projections, hence the
number of helicity states for given values of $J$ and $j'$. Note that as the initial
rotational state is $j$=0, there is only one parity and the number of helicities, $\Omega'$,
is $[\min(J,j')+1]$ and not $[2\min(J,j')+1]$ as in those cases where  two triatomic
parities are involved. Following Eq.~\eqref{sm}, the statistical prediction at
$10^{-7}$\,K, for which there is only a single projection, $\Omega'$=0,  is that all
rovibrational states would exhibit the same cross section. However, although all
rotational states are populated, as shown in Fig.~\ref{fig6}, there are significant
deviations from this expectation. Even if we leave aside the presence of peaks at $j'$=9,
11 for $v'$=0, and $j'$=4 for $v'$=1, whose origin has to be dynamical, the rotational
states for $v'$=1 are more populated. This is an indication of the preference for those
states with the largest content of internal energy. The internal energy of the rotational
states of $v'$=1 is only comparable to those of highest rotational states of $v'$=0
populated ($j'$=19, 20) and, as mentioned, they should exhibit similar cross sections.
However, when $J$=0, $l'=j'$ and therefore highly excited rotational states, whose
available translational energy is very small, will be subject to the highest centrifugal
barriers in the exit channel.  Consequently,  the rotational states of the $v'$=1 manifold
display a larger cross section. It is clear that in the ultracold regime, the reaction cross
sections are dominated by dynamical features (oscillations, sharp peaks) that stand out
against the usual statistical pattern  that can be found at higher energies.

Indeed, apart from some conspicuous oscillations (minima and maxima) whose
origin is doubtless dynamical,  the rotational distributions at 10\,K seem to
conform better with the statistical behavior that can be found at thermal and
hyperthermal energies.  At this energy $J_{\max} \approx 11$,  giving rise to
many more projections, whose number increases with $j'$ until $j'$=11.  For  $v'$=0, 1
the behavior  follows the statistical pattern:  the cross section increases
with $(j'+1)$ until the limiting value energetically allowed. In both cases
the slope resulting from plotting the ratio
$\sigma(v',j')/\sigma(v',0)$ vs. $(j'+1)$ for $j' <$11 is close to one.

At 220\,K , where many partial waves contribute to all the states, the behavior
is essentially statistical. The cross section is proportional to $(j'+1)$ up to
$j'$=15 for $v'$=0. The cross sections of $j'$ states with $v'$=1
become closer to the respective rotational states with $v'$=0.
\begin{figure}[h!]
 \begin{center}
 \hspace{-1cm}
 \vspace{-0cm}
\includegraphics[width=60ex]{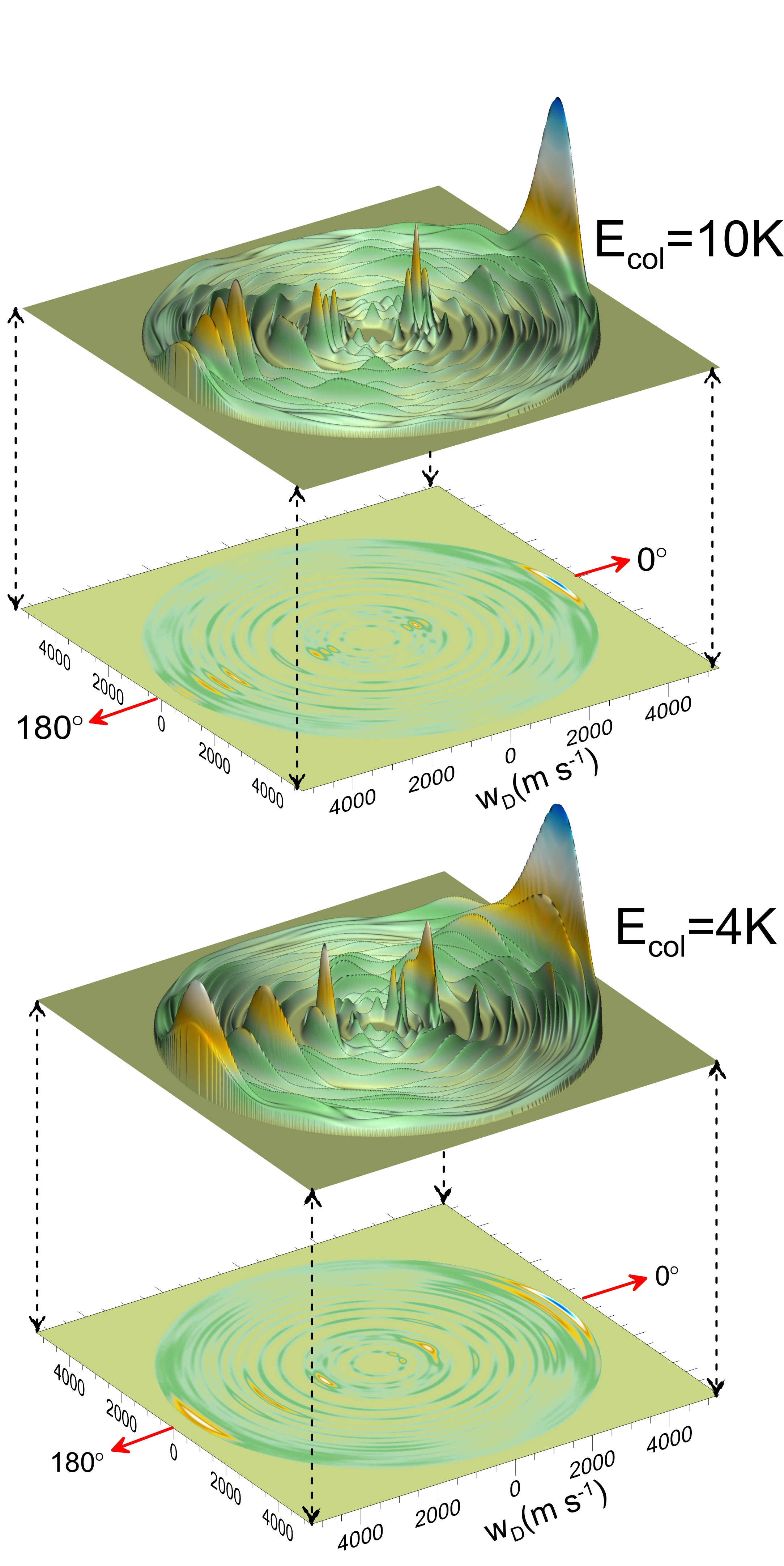}
\caption{Same as in Fig.\ref{fig7} at the collision energies of 10\,K (upper panel) and 4\,K
(bottom panel). In both cases, there is a predominance for forward scattering.
Note that the scattering angle is defined as the angle between the recoil direction of the SD moiety
and that of the incoming S atom.  The inner rings correspond to the $v'$=1 manifold.}
 \label{fig8}
\end{center}
\vspace{-0.5cm}
\end{figure}

The main conclusion of the observed rovibrational distributions is that the
dynamical effects prevail over the statistical behavior at collision energies
where only few partial waves are open ($<10$\,K). In the ultracold regime, when
only the s-partial wave comes about, the effect is even more striking since the
statistical prediction, apart from the centrifugal barrier in the exit channel,
would assign the same weight to all states, $[\min(J=0, j')+1]=1$. Of course, the
specific features that show up in the calculations are dependent on the PES
and subject to its accuracy, therefore should be taken with
caution.

The double differential cross sections (DCS), that is, the joint angular and
recoil velocity distributions of the D-atom product, are portrayed in
Figs.~\ref{fig7} and \ref{fig8}.  These polar maps would be similar to those
that could be observed experimentally (for instance, using D-atom Rydberg
tagging or VMI after ionization of D)  if the angle-recoil velocity of the
emerging D-atom were observed with a resolution of $\approx$120 m s$^{-1}$. The
DCS at 10$^{-7}$\,K, shown in the upper panel of Fig.~\ref{fig7}, is isotropic
since only $J$=0 contributes to the scattering.  The distribution of the D-atom
center-of-mass  recoil velocity is shown in the lower panel of Fig.~\ref{fig7}
and corresponds to the rovibrational distribution shown in Fig~\ref{fig6}. As
could be expected, the peaks of the $v'$=1 rotational states are considerably
more pronounced than those from $v'$=0.

The polar maps at 10\,K and 4\,K  are shown in Fig.~\ref{fig8}. They are no longer
isotropic, with some preponderance of forward scattering (the direction of the incoming
S($^1$D) atom).  A  strict statistical behavior would have implied a forward-backward
symmetry in DCS.  However, a lack of symmetry can be also expected within the
statistical ansatz, as it has been discussed previously for the S($^1$D)+H$_2$ reaction
, and previous calculations have shown that the
preference towards forward or backward scattering varies markedly with small changes
in the collision energy.\cite{Bonnet:JCP2015,JLA:JCP21}  Although there is not vibrational population inversion, the
sharp peaks associated to rotational states of $v'$=1 (inner rings) are conspicuous.

\section{Summary and Conclusions}

Fully converged, nearly exact adiabatic quantum mechanical calculations have
been performed for elastic and reactive  S($^1$D)+H$_2(v=0,j=0)$ collisions at
energies, $E/k_BT$,  between  10$^{-8}$\,K and  220\,K, covering the
ranges of ultracold, cold and thermal regimes.

Since the reaction is barrierless and complex-mediated, the Langevin model is likely to
provide a rough estimate of the QM reaction cross section in the thermal and
upper-cold regime. In this work, the classical Langevin model has been improved using
a numerical capture statistical model that includes discretization of the angular
momentum and uses the actual heights of the effective potential in the entrance
channel. It is shown, however, that the Langevin model is not applicable in the ultracold
regime since it  does not fulfil the limiting Wigner threshold behavior.

The universal system, as defined in the Multi-Quantum-Defect theory (MQDT),
constitutes  a good model for complex-mediated systems where the number of product
states is so high that we can assume that most of the probability is irreversibly
dissipated from the incoming channel into  the well.  The agreement between the QM
results and the results from the universal model is satisfactory (factors lower than 2 in
terms of partial cross-sections) for low partial waves and on the same order of
magnitude than those obtained in the study of the D$^+$ +H$_2$ system,  where the QM
results were compared with those  obtained  by the MQDT expressions for a
non-universal case. Interestingly, it is found that the reaction probability in the
ultracold limit is higher than the capture value. The paradox is partially resolved
bearing in mind that there is no quantum condition for a perfect quantum absorption,
and we are simply finding the solution using an approximate  semiclassical boundary
condition.

State-to-state integral and differential cross sections are also shown at several collision
energies. The QM results indicate that the statistical behavior of the rovibrational
distributions is only partially fulfilled below 10\,K. In particular, the calculations show
that there is a vibrational population inversion in the ultracold regime, where only
$J$=0 contributes.  Based on statistical grounds, when $J$=0 is the only partial wave
($E<10^{-2}$\,K) we can expect that all the rovibrational states are equally
populated if the effect of the centrifugal barrier is disregarded. However, under these
conditions it seems that dynamical effects outweigh the statistical behavior. By
increasing the collision energy, the state resolved integral cross sections follow  the
expected pattern characteristic of statistical behavior, as demonstrated by the fact that
the final-state resolved cross sections are proportional to $(j'+1)$.  Although the differential cross
sections at ultracold energies are necessarily isotropic, the velocity distribution reflects
the population inversion of the SD products. As the collision energy increases, the
angular distributions are not isotropic any longer, and exhibit the expected
forward-backward symmetry characteristic of statistical reactions albeit with some
preponderance for forward scattering.

\section{Appendix}

\subsection{The Multi-Quantum Defect theory (MQDT) model} \label{A1}

The formalism in Ref.~\cite{Jach:02} provides expressions for the complex (energy
dependent) scattering length, $\tilde{a}_{l}(k)=\alpha_{l}(k)-i \beta_{l}(k)$ in terms of
the MQDT functions. This allows to parameterize $\tilde{a}_{l}(k)$ using two real
parameters, $y$ and $s$, together with the mean scattering length, $\bar{a}$, given by
Eq.~\eqref{mean_a}. Specifically, the parameter $0 \le y \le 1$ characterizes the {\em
loss probability at short range}, {\em i.e.} the
 flux which is lost from the incoming channel at SR, $P^{\rm re}$,
according to $P^{\rm re}=4y/(1+y)^2$. The dimensionless scattering length,
${s=\tan\phi}$, is related to an entrance channel phase $\phi$.
\cite{Jach:02,Jach:03,Simoni}. In terms of these parameters, and following
Ref.~\citenum{Jach:03}, the small $k$ behavior of the complex scattering length for the
lowest partial waves ($l=0, 1$) in the case $n$=6 is given by:
\begin{align}
\alpha_{0}(k)&\to  \bar a \, \frac{s+y^2(1-s)(2-s)}{1+y^2 (1-s)^2} , \label{una}\\
 \beta_{0}(k) &\to  y \,\frac{1+(1-s)^2}{1+y^2 (1-s)^2} \, \bar a , \label{unap} \\
\alpha_{1}(k) &\to  -2 \bar{a}_1 \, (k \bar{a})^2 \, \frac{y^2s+(s-1)(s-2)}{y^2s^2+(s-2)^2},
\label{dos} \\
\beta_{1}(k) &\to  -2  \bar{a}_1 \, (k \bar{a})^2 \,\frac{y(2s-s^2-2)}{y^2s^2+(s-2)^2} ,
\label{dosp}
\end{align}
where \begin{equation}\label{a1bar}
 \bar{a}_1=\bar{a}\,\Gamma(1/4)^6/(144 \pi^2 /
\Gamma(3/4)^2) \approx 1.064\,\bar{a}\,.
\end{equation}

A {\em quantum Langevin} behavior or {\em universal case} would correspond to $y=1$.
Particularizing expressions ~\eqref{una}-\eqref{dos} to this case we get:
\begin{align}
\alpha_{0}(k) &\to  \bar a,
\quad \beta_{0}(k) \to \bar a
\label{unab} \\
\alpha_{1}(k)  &\to  -\bar{a}_1 \, (k \bar{a})^2\,, \qquad
\beta_{1}(k) \to  \bar{a}_1 \, (k \bar{a})^2
\label{dosb}
\end{align}
Using Eqs.\eqref{losess0} and \eqref{losess}, we can get the values of the
elastic and reaction cross-sections for the two lower partial
waves:
\begin{align}
  \sigma^{00}_{\rm r} (E) &\to  4 \pi  \bar a
  \left( \frac{\hbar^2}{2 \mu E} \right)^{1/2}\,,\label{unot}\\
 \sigma^{00}_{\rm el} (E)&\to 8 \pi  {\bar a}^2\,,
\label{dost} \\
\sigma^{1m}_{\rm r} (E)&\to  4 \pi  \bar{a}_1 {\bar a}^2
  \left( \frac{2 \mu E}{\hbar^2} \right)^{1/2}\,, \label{trest}\\
\sigma^{1m}_{\rm el} (E)&\to
8 \pi (\bar{a}_1 {\bar a}^2)^2 \left(\frac{2 \mu E}{\hbar^2}\right)^2
\label{cuatrot}
\end{align}

\subsection{Calculating capture probabilities}\label{A2}
Following the approach of the MQDT model \cite{Jach:02}, we can assume that, as the
collision partners approach, the incoming probability flux remains associated to the incoming
channel, submitted to the $V_{LR}(R)=-C_6/R^{-6}$ LR potential. The 1D radial TISE (in
reduced form) corresponding to that channel is given by
\begin{eqnarray}\label{eq:acoplamien}
 \left[ - \frac{\hbar^2}{2 \mu}  \frac{\partial^2}{\partial
R^2}  + \frac{l(l+1) \hbar^2 }{2 \mu R^2} +V_{LR}(R)\right]\psi(R) =  E \psi(R),
\end{eqnarray}
Only at short distances, lower that a hypothetical $R_0$, the probability is transferred to
other collision channels. Assuming the validity of the WKB approximation in an intermediate
region ($R_0 < R <<R_6$), we can distinguish incoming and outgoing terms in the solution
to the equation, and express it as:
\begin{eqnarray}
  \psi(R) \sim \frac{e^{- i \int_{R} k(x) dx}}{\sqrt{k(R)}} - (1-P^{\rm re})^{1/2}e^{i \phi
  }\frac{e^{+ i \int_{R} k(x) dx}}{\sqrt{k(R)}}
\end{eqnarray}
where $k(x)=\sqrt{2 \mu (E-V_{\rm LR}(x))}/ \hbar$. The first term represents the incoming
probability flux, and the second one the reflected one. If $P^{\rm re}$ notes the probability
which is irreversibly lost from the incoming channel at short range, the reflected probability is
$(1-P^{\rm re}). $\footnote{for the sake of clarity, we prefer to introduce the reflected
amplitude as $(1-P^{\rm re})^{1/2}$,   instead of the expression $(1-y)/(1+y)$ used in
Ref.~\citenum{Jach:02}.  Given the relation $P^{\rm re}=4y/(1+y)^2$, both are equivalent}
Hence we can write the amplitude of the reflected term as $(1-P^{\rm re})^{1/2}e^{i \phi }$,
where $\phi$ accounts for any possible phase difference between both. The real and
imaginary parts of the logaritmic derivative at a particular point of the considered region,
$Z(R)=\psi'(R)/\psi(R)$, are given by:
\begin{eqnarray}\label{zeta}
Re(Z)&=& -\frac{k}{2k'}   \nonumber \\
Im(Z)&=&-k\left[\frac{P^{\rm re}+2(1-P^{\rm re})^{1/2}\sin\phi}{2-P^{\rm re}-2(1-P^{\rm
  re})^{1/2}\cos\phi}  \right]
\end{eqnarray}
The capture probability corresponds to the case $P^{\rm re}=1$, which means a perfect
absorption (no reflection term) at SR. The logarithmic derivative takes then a very
simplified form $Z(R)=-ik$ (the real part can be neglected if the WKB approximation is
fulfilled). We can use the method by De Vogelaere \cite{Lester} to find a solution with such a
logarithmic derivative at a particular matching point $R_m$. We proceed as follows. By
inwards integration, we obtain the ``regular'' ($F^{(1)}_{l}$) and ``irregular'' ($F^{(2)}_{l}$)
solutions, defined as the ones  which behave as
 \begin{align}\label{eq:desar2}
F^{(1)}_{ l}(R)& \stackrel{R >> R_6} {\longrightarrow} k^{1/2} R\, \,j_l(kR)
           \stackrel{R \to \infty} {\longrightarrow}  \frac{\sin(kR-l \pi /2)}{k^{1/2}},  \nonumber \\
           F^{(2)}_{l} (R)& \stackrel{R >> R_6} {\longrightarrow} k^{1/2} R \,\,n_l(kR)
           \stackrel{R \to \infty} {\longrightarrow} -\frac{\cos(kR-l \pi /2)}{k^{1/2}},
\end{align}
where $j_l$ and $n_l$ are the regular and irregular spherical Bessel functions. As usual,
we assume that the asymptotic behavior of the wavefunction is given by
 ${\psi(r) \sim k^{1/2} R [ j_l(kR) - \tan (\delta_l) n_l(kR) ]}$ at long distance, $\delta_l$ being
 the phase shift, and hence by
 ${\psi(r) \sim F^{(1)}_{ l} -\tan (\delta_l)  F^{(2)}}$ at shorter radius. In particular, at a point
 $R_m$  (where the WKB
 approximation is fulfilled) we equate $\psi'(r)/\psi(r)$ to the perfect absorption logarithmic
 derivative, $Z(R)=-ik$, and solve for
 the value of $\tan (\delta_l)$, also called reactance matrix $K$.  We obtain
 \begin{equation}
 \!\!\!K(E)= \tan \delta_l (E) =\frac
   {F^{(1)}_{ l}(R_m)-Z(R_m)F'^{(1)}_{ l}(R_m)}
   {F^{(2)}_{ l}(R_m)-Z(R_m)F'^{(2)}_{ l}(R_m)}
 \end{equation}
And the value of the $S$-matrix is given by
 \begin{eqnarray}
 S(E)=e^{2i\delta_l}=\frac{1+iK(E)}{1-iK(E)}
 \end{eqnarray}
The obtained S-matrix element will not have modulus 1, due to the loss of flux at short
distances. The lack of unitarity, $1-|S|^2$ is precisely the capture probability. Let us finally
note that partial absorptions and the corresponding S-matrix elements can be simulated by matching to
any other value of the logarithmic derivative given by Eq. \ref{zeta}.

\subsection{Calculation of the real part of the $l$-dependent complex scattering length in the
case of anomalous behavior}

In the case of a LR interaction of the type $V_{\rm LR}(R)=-C_6/R^{-6}$, and for
values of $l>1$, the phase shift is dominated by a term $\sim E^2$. We talk of
anomalous behavior. The main contribution to the phase shift is due to the LR part of
the potential. It is precisely in this region where the potential is weak enough for the
Born approximation to work and we can use it to calculate the phase shift. Following Chapter V,
section 2, of Ref. \cite{Mott}, the phase shift as a function of $l$ is given by
\begin{widetext}
 \begin{eqnarray}
  \delta_l=\frac{3\pi \mu C_6 k^4}{16 \hbar^2(l+5/2)(l+3/2)(l+1/2)(l-1/2)(l-3/2)}
 \end{eqnarray}
 Using this expression, the real part of the scattering length is
  \begin{align}
  \alpha_l = -\frac{1}{k}\tan\delta_l   \stackrel{E \rightarrow 0} {\longrightarrow} -\frac{3 \pi
  \mu C_6 k^3}{16
  \hbar^2(l+5/2)(l+3/2)(l+1/2)(l-1/2)(l-3/2)} \label{elasalfa}
 \end{align}
\end{widetext}

\section{acknowledgments}
M. L. is very grateful to J.-M. Launay for his kind help and support, to Andrea Simoni for
his insightful discussions and helpful comments, and to K. Jachymski for his willingness 
to provide details of on his MQDT theory.
M. L.
and  F.J.A. acknowledge funding by the
Spanish Ministry of Science and Innovation (Grants No. PGC2018-096444-B-I00 and
PID2021-122839NB-I00).  P.G.J.  acknowledges grant
PID2020-113147GA-I00 funded by MCIN/AEI/10.13039/501100011033.

\bibliographystyle{apsrev4-1}

%

\end{document}